\documentclass[runningheads]{llncs}
\usepackage{graphicx}
\usepackage{booktabs}
\usepackage{amsmath}
\usepackage{amsfonts}
\usepackage{url}
 \usepackage[misc]{ifsym}
\newcommand{\vect}[1]{\mathbf{#1}}

\usepackage[symbol]{footmisc}

\begin{document}
\title{Distributional Correlation--Aware Knowledge Distillation for Stock Trading Volume Prediction}

\author{Lei Li\inst{1,2}, Zhiyuan Zhang\inst{1,2}, Ruihan Bao\inst{3}\Letter, Keiko Harimoto\inst{3}, Xu Sun\inst{1,2}\Letter}

\institute{MOE Key Lab of Computational Linguistics, Peking University \and School of Computer Science, Peking University \and Mizuho Securities Co., Ltd \\ 
\email{lilei@stu.pku.edu.cn} \quad \email{\{zzy1210, xusun\}@pku.edu.cn} \\
\email{\{ruihan.bao, keiko.harimoto\}@mizuho-sc.com}
}
\tocauthor{}
\toctitle{}
% First Author\inst{1}\orcidID{0000-1111-2222-3333} \and

\maketitle              % typeset the header of the contribution
\begin{abstract}

Traditional knowledge distillation in classification problems transfers the knowledge via class correlations in the soft label produced by teacher models, which are not available in regression problems like stock trading volume prediction.
To remedy this, we present a novel distillation framework for training a light-weight student model to perform trading volume prediction given historical transaction data.
Specifically, we turn the regression model into a probabilistic forecasting model, by training models to predict a Gaussian distribution to which the trading volume belongs.
The student model can thus learn from the teacher at a more informative distributional level, by matching its predicted distributions to that of the teacher.
Two correlational distillation objectives are further introduced to encourage the student to produce consistent pair-wise relationships with the teacher model.
We evaluate the framework on a real-world stock volume dataset with two different time window settings. Experiments demonstrate that our framework is superior to strong baseline models, compressing the model size by $5\times$ while maintaining $99.6\%$ prediction accuracy. The extensive analysis further reveals that our framework is more effective than vanilla distillation methods under low-resource scenarios.\footnote{Our code and data are available at \url{https://github.com/lancopku/DCKD}.}

\end{abstract}
\section{Introduction}
Large deep neural networks~(DNNs) like Transformer~\cite{vaswani2017transformer} have achieved superior performance in various areas like computer vision~\cite{dosovitskiy2020vit}, natural language processing~\cite{devlin2019bert} and time series forecasting problems like stock trading volume prediction~\cite{zhang2021asat}.
However, the increase of model parameters demands more computational resources, limiting their applicability in latency-sensitive scenarios like high-frequency trading~(HFT). 
The pursuit of a better performance-efficiency trade-off promotes an active research field toward compressing large DNNs while maintaining promising model performance.
% Progresses have been made towards compressing DNNs in the recent years. 
Pilot model compression techniques include pruning~\cite{han2015deep}, quantization~\cite{hubara2016binaryNN,Shen2020QBERT} and knowledge distillation~\cite{Hinton2015Distilling,romero2014fitnets}.
Pruning improves the parameter efficiency by de-activating redundant structures in the network, and quantization focuses on exploring fewer bits for representing the model weights. 
While effective in reducing the model size, these two methods require hardware-specific support for actually gaining the speed-up.
% which removes redundant structures in the network and quantization~\cite{hubara2016binaryNN,Shen2020QBERT} which aims to represent the weight with fewer bits, have been prove effective in reducing the model parameters and maintaining superior performance, while both requiring hardware-specific support for actually gaining the speed-up.
% 讲一下 KD
On the other hand, knowledge distillation~(KD), trains a much smaller student model by utilizing the learned knowledge from a large teacher model. It has been prove successful in various classifications problems like natural language understanding~\cite{Sanh2019DistilBERT,Sun2019PatientKD} and image classification~\cite{Hinton2015Distilling,romero2014fitnets}, and 
recent studies have demonstrated that KD can obtain a compact student model that matches or even outperforms the teacher model~\cite{romero2014fitnets,furlanello2018born,Jiao2019TinyBERT}.

Traditional knowledge distillation works relatively well for classification problems, as it can transfer the \emph{dark knowledge}, i.e., the softened logits of the teacher prediction, to the student. 
The softened logits contain richer supervision signals than the vanilla one-hot class label, reflecting the semantic correlation between different classes and thus boosting student performance.
% can be boosted, producing a better performance-efficiency trade-off.
However, this advantage cannot hold in regression problems like stock trading volume prediction, as the teacher model only produces real-valued predictions which have an identical characteristic to the oracle label. 
Without an optimal carrier for the learned knowledge in the teacher, the effects of KD are limited in regression problems.

\begin{figure}[tbh!]
    \centering
    \caption{Distributional knowledge distillation for regression problems. 
    While conventional KD for regression problems only transfers the knowledge by matching the scalar $y_S$ to $y_T$, our proposed distributional KD operates on the distributional level and provides more informative supervision for the student.}
    \includegraphics[width=0.9\linewidth]{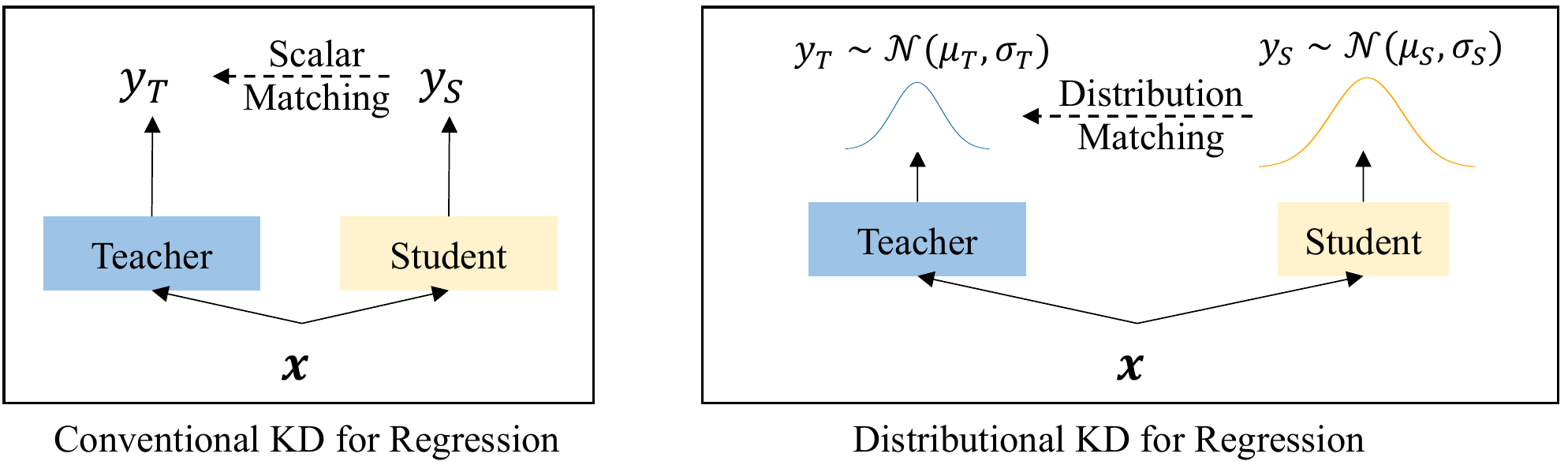}
    \label{fig:dkd_overview}
\end{figure}
% 另外一点 是不是需要一个更金融的建模方案呢? 
% While conventional KD transfers individual outputs from a teacher
% model (fT ) to a student model (fS) point-wise, our approach transfers relations of the outputs structure-wise. It
% can be viewed as a generalization of conventional KD.

% unbounded real-valued predictions.
% predicts
% sequential, continuous, values which have the exact same
% characteristics as the ground truth,
% , which are not informative for the student. Worsenly, erroneous teacher predictions can even mis-guidance the student.
% The single value prediction is not a optimal carrier of the learned dark knowledge.

To remedy this, in this paper, we propose a distributional knowledge distillation framework for regression problems, as illustrated in Figure~\ref{fig:dkd_overview}.
Specifically, we first turn the problem into a probabilistic forecasting problem. We cast the trading volume prediction problem as a conditional probability distribution modeling problem given the historical data.
The teacher and the student are both probabilistic forecasting models trained by minimizing the log-likelihood of the training data.
The learned knowledge from the teacher model is then transferred to the student by minimizing the discrepancy between the predicted distributions. 
% Besides, a the capacity gap poses challenges for the student model to mimicking the predictions of the teacher model, 
% Furthermore, in order to make the student model predict more consistently with the teacher model and alleviate t
Besides, recent studies have shown that the capacity gap between the teacher model and the student model may harm the distillation effect~\cite{Mirzadeh2020TAKD,Li2021DynamicKD}, which is also observed in our vanilla distributional KD framework.
To alleviate this, we design two correlations between different samples regarding the output distributions.  
% that capture the mutual relationship  in the mini-batch. 
The student is then trained to predict pair-wise correlations consistently with the teacher model, by incorporating the correlation congruence objectives into the distillation process. 
These objectives serve as auxiliary objectives to provide more informative supervision for alleviating the capacity gap problem.
We validate our proposal by distilling a multiple-layer Transformer model into a single layer student model.
Experiments on a real-world stock volume prediction dataset show that our framework can reduce the number of model parameters by $5$ times while maintaining $99.6\%$ prediction accuracy.
Further analysis shows that our framework is more effective under low-resource settings and can make the student produce more calibrated predictions.

\section{Methodology}
In this section, we first formulate the stock trading volume prediction problem and introduce the metrics defined for evaluation.
Followingly, we introduce conventional knowledge distillation for classification problems. We then elaborate the proposed distributional correlation-aware knowledge distillation framework for the regression problem. 
% Our framework first converts the regression problem into a probability forecasting modeling problem. Then the learned knowledge of the teacher model is transferred into the student model via designed correlation-aware distillation objectives.

% Given a series of hourly transaction
% data points in the past few days containing N data points
% X = {x1, x2, . . . , xN }, of which each xi consists of volume data x
% p
% i
% and price data x
% v
% i
% . Each x
% p
% consists of hourly
% highest price, lowest price, open price and close price. Each
% x
% v
% consists of hourly volume v and the proportion of it in the
% whole day. Besides, there are some events encoded as news
% information D released in the market.
\subsection{Task Formulation}
Stock trading volume prediction aims at predicting the market trading volume given historical transaction data. Specifically, given training dataset consists of $N$ data samples $\mathcal{D} = \{ (\mathbf{x}_1, y_1), \ldots, (\mathbf{x}_N, y_N) \}$, where $\mathbf{x}_i$ denotes the transaction data including open, closing, lowest, highest price and the trading volume in the past time windows, and $y_i$ is the target volume of the $i$-th sample.
Our goal is training a light-weight student model $S$, to predict the trading volume $\hat y$, by learning from a larger teacher model $T$.
The student performance is measured by the mean squared error~(MSE), mean absolute error~(MAE) and prediction accuracy~(ACC):
\begin{equation}
\begin{aligned}
\mathrm{MSE} &=\mathbb{E}_{(\mathbf{x}, y) \sim \mathcal{D}}(\hat{y}-y)^{2}, \\
% \mathrm{RMSE} &=\sqrt{\mathbb{E}_{(\mathbf{x}, y) \sim \mathcal{D}}(\hat{y}-y)^{2}} \\
\mathrm{MAE} &=\mathbb{E}_{(\mathbf{x}, y) \sim \mathcal{D}}|\hat{y}-y|, \\
\mathrm{ACC} &=\mathbb{P}_{(\mathbf{x}, y) \sim \mathcal{D}}\left(\left(\hat{y}-y_{\text {last }}\right) \times\left(y-y_{\text {last }}\right)>0\right),
\end{aligned}
\end{equation}
where $y_\text{last}$ is the volume of the most last time slot. Thus, ACC is the accuracy of whether the volume increases or decreases compared to the last time slot.

\subsection{Conventional Knowledge Distillation for Classification}
Knowledge distillation is a classic framework for transferring the knowledge of a larger teacher model to a light-weight student model. The main idea behind is training the student model to mimic the outputs of the teacher model.
Specifically, in a classification problem, given the one-hot label $\mathbf{y}$, the student prediction $\mathbf{O}_S$ and the teacher prediction $\mathbf{O}_T$ over the class set, KD is usually achieved by minimizing both the hard label error and a soft label error between the student and the teacher predictions:
% and KL-divergence between the output class distributions:
\begin{equation}
\mathcal{L}_{KD}=\alpha \mathcal{H}\left(\mathbf{y}, \mathbf{O}_{S}\right)+(1-\alpha) \mathcal{H}\left(\mathbf{O}_{T}, \mathbf{O}_{S}\right),
\end{equation}
where $\mathcal{H}(\cdot, \cdot)$ denotes the cross-entropy objective and $\alpha$ is a tuning parameter controlling the relative contribution of cross-entropies.
As $\mathbf{O}_T$ usually contains rich information regarding the semantic relationships between classes, the student can capture more fine-grained structured information from the teacher predictions than directly learning from the ground-truth label.
However, this characteristic cannot hold in regression problems like stock trading volume prediction, as the teacher predictions are also real-valued scalars. The predicted scalars of the teacher cannot convey more information to benefit the student, motivating us to explore a better distillation framework for regression problems.

\subsection{Distributional Knowledge Distillation for Regression Problems}
To facilitate the distillation effect, we propose to cast the trading volume prediction problem as a probabilistic forecasting problem, thus the information can be transferred at the distribution level.
Specifically, following DeepAR~\cite{salinas2020deepar}, instead of directly predicting the scalar, we assume that the predicted trading volume follows a Gaussian distribution $\mathcal{N}(\mu,\sigma)$, turning the regression problem into a likelihood model as:
% and model the mean $\mu$ and the variance  $\sigma$ of the Gaussian distribution.
\begin{equation}
\begin{aligned}
%p(y  \mid \mu, \sigma) &=\left(2 \pi \sigma^{2}\right)^{-\frac{1}{2}} \exp \left(-(y-\mu)^{2} /\left(2 \sigma^{2}\right)\right)
p(y  \mid \mu, \sigma) &=\frac{1}{\sqrt{2 \pi \sigma^{2}}}\exp\left(-\frac{(y-\mu)^{2}}{2 \sigma^{2}}\right).
\end{aligned}
\end{equation}
The Gaussian distribution is parameterized by a mean $\mu$ and a standard deviation $\sigma$, which can be obtained by applying affine transformations on the model encoding $\mathbf{h}$ of the input transaction data $\mathbf{x}$:
\begin{equation}
\begin{aligned}
\mu\left(\mathbf{h} \right) &=\mathbf{w}_{\mu}^{T} \mathbf{h}+b_{\mu} \\
\sigma\left(\mathbf{h} \right) &=\log \left(1+\exp \left(\mathbf{w}_{\sigma}^{T} \mathbf{h} +b_{\sigma}\right)\right),
\end{aligned}
\end{equation}
where $\mathbf{w}_{\mu}$, $b_{\mu}$, $\mathbf{w}_{\sigma}$ and $b_{\sigma}$ are learnable parameters of the affine transformation.
Note that the standard deviation is wrapped with a softplus activation to ensure the value is positive.
With this formulation, a model $M$ can be trained by minimizing the negative log-likelihood of the ground-truth data:

\begin{equation}
\begin{aligned}
\mathbf{h}_{i} &= M( \mathbf{x}_i) \\ 
\mathcal{L}_\text{NLL} &=- \sum_{i=1}^{N} \log p \left(y_i \mid \mu\left(\mathbf{h}_{i}\right), \sigma\left(\mathbf{h}_i\right) \right).
\end{aligned}
\end{equation}
We first train a teacher model with the above objective, and then transfer the learned knowledge into the student model by minimizing the Kullback-Leibler~(KL) divergence between the Gaussian distributions~\cite{pardo2018gaussianKL} predicted by the teacher and the student model:
% log ( sigma_2 / sigma_1) + \frac{ sigma_1 ** 2  + (mu_1 - mu_2) ** 2 }{2 * sigma_2 ** 2} - 0.5  

\begin{equation}
\begin{aligned}
\mathcal{L}_{\text{DKD}} &= - \sum_i^N \text{KL}\left(  \mathcal{N}\left(\mu^{T}_i, \sigma^{T}_i\right) \|   \mathcal{N}\left(\mu^{S}_i, \sigma^{S}_i\right)  \right) \\
&= - \sum_i^N \left( \log \frac{\sigma^S_i}{\sigma^T_i} + \frac{ \left(\sigma^T_i\right)^2 + \left(\mu^T_i - \mu^S_i \right)^2}{2\left(\sigma^S_i\right)^2} - \frac{1}{2} \right) , \\ 
% \mathcal{L}_{S} &= \mathcal{L}_{\text{DKD}} 
\end{aligned}
\label{eq:dkd}
\end{equation}
where $\mu^S_i$, $\mu^T_i$, $\sigma^S_i$ and $\sigma^T_i$ are the mean and standard deviation outputs of the student model and the teacher model of the $i$-th data sample, respectively.
% \begin{equation}
% \begin{aligned}
% \mathcal{L}_{DKD} = KL(p, q) &=-\int p(x) \log q(x) d x+\int p(x) \log p(x) d x \\
% &=\frac{1}{2} \log \left(2 \pi \sigma_{2}^{2}\right)+\frac{\sigma_{1}^{2}+\left(\mu_{1}-\mu_{2}\right)^{2}}{2 \sigma_{2}^{2}}-\frac{1}{2}\left(1+\log 2 \pi \sigma_{1}^{2}\right) \\
% &=\log \frac{\sigma_{2}}{\sigma_{1}}+\frac{\sigma_{1}^{2}+\left(\mu_{1}-\mu_{2}\right)^{2}}{2 \sigma_{2}^{2}} .
% \end{aligned}
% \end{equation}
\subsection{Transferring Knowledge via Correlation Consistency}
Directly minimizing the KL-divergence between distributions can be challenging for the student model, as revealed by recent studies regarding the capacity gap between the teacher model and the student model~\cite{Mirzadeh2020TAKD,Li2021DynamicKD}.
To remedy this, we introduce correlational knowledge distillation objectives which capture the pair-wise relationships between the examples for alleviating this issue.
Specifically, given the outputs distributions of the teacher models and the students model on $m$ data samples:
\begin{equation}
\begin{aligned}
    \boldsymbol{N}_T &= \left[\mathcal{N}_1^T, \ldots, \mathcal{N}_m^T \right] \\ 
    \boldsymbol{N}_S &=  \left[\mathcal{N}_1^S, \ldots, \mathcal{N}_m^S \right].
\end{aligned}
\end{equation}
A mapping function $\psi$ is
introduced for mapping the outputs to a pairwise correlation matrix $\boldsymbol{C}$:
\begin{equation}
\psi: \boldsymbol{N} \rightarrow \boldsymbol{C} \in \mathbb{R}^{m \times m}.
\end{equation}
The element in $\boldsymbol{C}$ denotes the correlation between distributions on two sample $\mathbf{x}_i$ and $\mathbf{x}_j$:
\begin{equation}
\boldsymbol{C}_{i j}=\varphi\left(\mathcal{N}_{i}, \mathcal{N}_{j}\right), \quad \boldsymbol{C}_{i j} \in \mathbb{R}.
\end{equation}
The function $\varphi$ denotes a correlation metric that captures the relationship between two Gaussian distributions, and the two options we designed for the function will be elaborated later. The correlational knowledge in the teacher then can be transferred by training student to minimize the congruence objective:
% The function ϕ can be any correlation metric, and we will
% introduce three metrics for capturing the correlation between instances in the next secti.
\begin{equation}
\begin{aligned}
\mathcal{L}_\text{DCKD} &=\frac{1}{m^{2}}\left\|\psi\left(\boldsymbol{N}_{S}\right)-\psi\left(\boldsymbol{N}_{T}\right)\right\|_{2}^{2} \\
&=\frac{1}{m^{2}} \sum_{i, j}\left(\varphi\left(\mathcal{N}_{i}^{S}, \mathcal{N}_{j}^{S}\right)-\varphi\left(\mathcal{N}_{i}^{T}, \mathcal{N}_{j}^{T}\right)\right)^{2}.
\end{aligned}
\end{equation}
In this way, the student can learn to predict the correlation between instances consistently with the teacher model.
The correlational distillation objective serves as an auxiliary objective. The student can first learn the correlations between its own predictions according to the teacher predictions, then make efforts towards predicting exactly the same as the teacher model.
% , thus mitigating the capacity gap problem and improving the distillation performance. 
Followingly, we introduce two correlation metrics regarding the distance-wise and the angle-wise similarity between samples.
% not only the instance level congruence but also correlation congruence between instances.

\subsubsection{Distance-wise Correlation}
Given two Gaussian distributions, a straightforward correlation metric is the distance between these two distributions.
Specifically, we adopt the Jeffreys divergence~(JSD) as the distance between $\mathcal{N}_i(\mu_i, \sigma_i)$ and $\mathcal{N}_j(\mu_j, \sigma_j)$ since it is symmetrized:
% 是否需要写一下证明 
% JF distance between two gaussin
% \mu_j, \sigma_j) \mu_i, \sigma_i)
\begin{equation}
\begin{aligned}
    \varphi_\text{Dist}(\mathcal{N}_i, \mathcal{N}_j) &= \frac{1}{2}\left(\text{KL}\left(\mathcal{N}_i \| \mathcal{N}_j\right) + \text{KL}\left(\mathcal{N}_j \| \mathcal{N}_i\right) \right)\\
    &=\frac{1}{2}  \left( \frac{1}{\sigma_i^2}  + \frac{1}{\sigma_j^2} \right)  \left(\left(\sigma_i - \sigma_j \right)^2 + \left(\mu_i - \mu_j\right)^2  \right).
\label{eq:dist}
\end{aligned}
\end{equation}
We note that the JSD can be replaced with alternative distance measurements like Wasserstein distance. As our goal is developing a distributional distillation framework for regression problems and the JSD works well in practice, we leave the explorations on the choices of distance metrics for future work.
\begin{figure}[t!]
    \centering
    \caption{The main idea illustration of the proposed distributional correlation-aware distillation, which transfers the knowledge in the teacher model by matching the predicted Gaussian distributions, with correlational consistency distillation objectives for improving the student performance.}
    \includegraphics[width=0.9\linewidth]{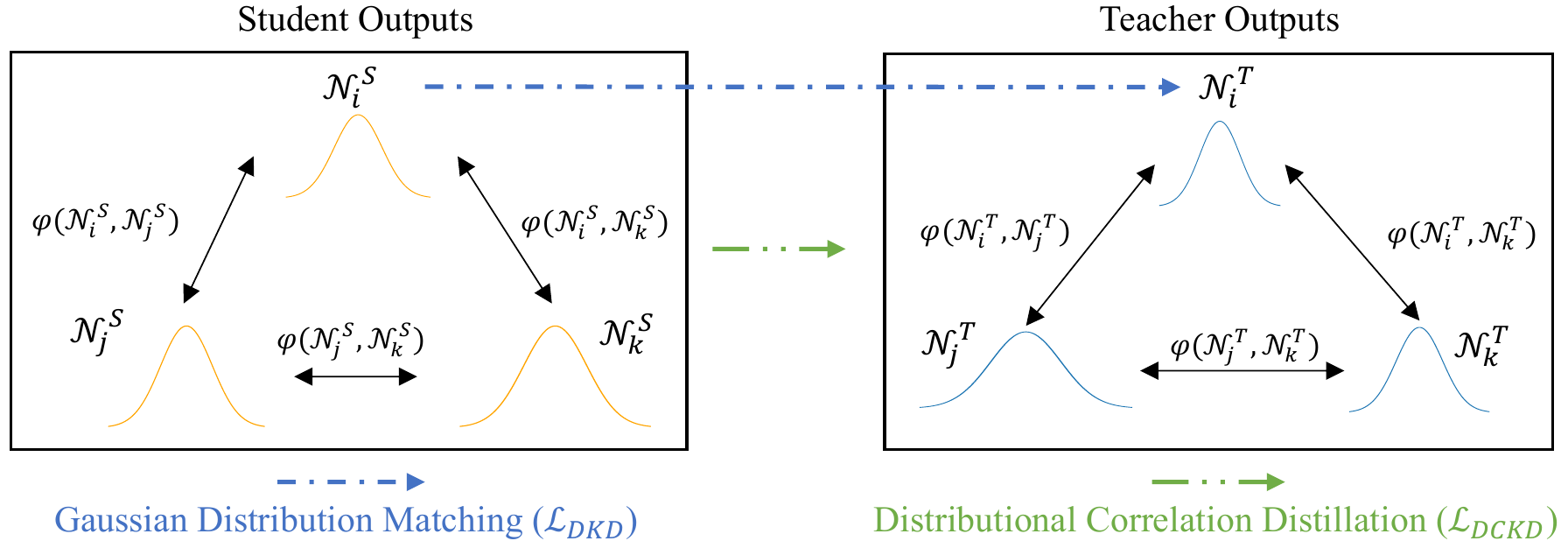}
    \label{fig:dckd}
\end{figure}

\subsubsection{Angle-wise Correlation}
%  Since an angle
% is a higher-order property than a distance, it may be able
% to transfer relational information more effectively, giving
% more flexibility to the student in trainin
Another commonly adopted similarity measurement is the angle-wise correlation, which is a higher-order relationship than the vanilla distance metric and thus can be more effective for transferring information~\cite{park2019relationalKD}.
In the Euclidean space, cosine similarity is a commonly adopted for evaluating the angle-wise correlation between two vectors:
\begin{equation}
\cos\langle\vect{u}, \vect{v}\rangle = \frac{(\vect{u}, \vect{v})}{\sqrt{(\vect{u}, \vect{u})(\vect{v}, \vect{v})}},
\end{equation}
where $\left(\vect{u}, \vect{v} \right)$ denotes the inner-product in the Euclidean space of two vectors. 
We extend this idea to Gaussian distributions and design a corresponding angle-wise cosine similarity metric for probabilistic distributions.
Specifically, we define $(\mathcal{N}_i, \mathcal{N}_j)$ as the inner-product in the Hilbert space:
% angle correlation, i.e., the cosine similarity between two Gaussian distributions as:
\begin{equation}
(\mathcal{N}_i, \mathcal{N}_j)=\int_{-\infty}^{+\infty}\mathcal{N}_i(t\mid\mu_i,\sigma_i)\mathcal{N}_j(t\mid\mu_j,\sigma_j)dt \  . 
\end{equation}
% Cosine
The cosine similarity thus can be calculated as:
% def cosine_pq(mu_1, sigma_1, mu_2, sigma_2):
%     return torch.exp( -(mu_1 - mu_2) ** 2 / ( 2 * (sigma_1 ** 2 + sigma_2 ** 2 ))) / torch.sqrt( ( (sigma_1 ** 2 + sigma_2 ** 2 ) / ( 2 * sigma_1 * sigma_2) ) + 1e-9) # (bsz, bsz)
\begin{equation}
    %\varphi_\text{Cosine}\left(\mathcal{N}_i, \mathcal{N}_j\right) = \frac{1}{\sqrt{\frac{\sigma_i^2 + \sigma_j^2}{2\sigma_i\sigma_j}}}\exp \left( - \frac{\left(\mu_i - \mu_j\right)^2}{2\left(\sigma_i^2 + \sigma_j^2\right)} \right)    
    \varphi_\text{Cosine}\left(\mathcal{N}_i, \mathcal{N}_j\right) =\frac{(\mathcal{N}_i, \mathcal{N}_j)}{\sqrt{(\mathcal{N}_i, \mathcal{N}_i)(\mathcal{N}_j, \mathcal{N}_j)}}= \sqrt{\frac{2\sigma_i\sigma_j}{{\sigma_i^2 + \sigma_j^2}}}\exp \left( - \frac{\left(\mu_i - \mu_j\right)^2}{2\left(\sigma_i^2 + \sigma_j^2\right)} \right).
\label{eq:cosine}
\end{equation}
We refer readers to Appendix~\ref{apx:proof} for the detailed proof of the inner-dot and the cosine similarity of Gaussian distributions.

By combining the correlation metrics with the distillation objective, the student model finally is trained by minimizing the following loss function:
\begin{equation}
    \mathcal{L} = \lambda_\text{NLL}\mathcal{L}_\text{NLL} + \lambda_\text{DKD}\mathcal{L}_\text{DKD} + \lambda_\text{Dist} \mathcal{L}_\text{Dist-CKD} + \lambda_\text{Cosine} \mathcal{L}_\text{Cosine-CKD},
\end{equation}
% \subsection{Training Objective}
where $\lambda_\text{NLL}$, $\lambda_\text{DKD}$,  $\lambda_\text{Dist}$ and $\lambda_\text{Cosine}$ are hyper-parameters for tuning the relative contribution of the proposed correlational distillation objectives.
We name the methods that setting $\lambda_\text{Dist} = 0$ and $\lambda_\text{Cosine} = 0$ as Cosine-CKD and Dist-CKD, respectively. 
Figure~\ref{fig:dckd} gives an overview of our proposed framework.

\section{Experiments}
In this section, we conduct experiments on a real-world stock trading volume prediction dataset for evaluating the effectiveness of our framework. We first introduce the dataset used for evaluation, followed by the details of compared baseline models and implementation details for reproducible results. 
Finally, we present the main results compared with strong baseline models.

\begin{table}[tbh]
	\centering
	\caption{The statistics of the TPX500 datasets used in our paper. The training and validation datasets have no time window overlapping with the test dataset to avoid potential data leakage.}
	\label{tab:dataset}
	\begin{tabular}{lcccccc}
		\toprule  
		\textbf{Dataset}&\multicolumn{3}{c}{\textbf{Hourly}}&\multicolumn{3}{c}{\textbf{Daily}}\\
		\midrule
		\textbf{Split}&Training & Validation &Test&Training &Validation&Test\\
		\midrule
		\textbf{\# of Samples}&  49,712 & 16,571 & 26,841& 81,950& 27,317& 38,316\\
		\bottomrule
		\end{tabular}
\end{table}

\subsection{Datasets}
We conduct our experiments by collecting trading data from the largest $500$ stock names traded at Tokyo Exchange known as TPX500.
For our research, we construct two datasets with different time windows, i.e., an hourly intra-day trading volume prediction dataset and a daily trading volume prediction dataset. 
The two datasets are both extracted from the price and trading volume data of the TPX500 between Jan. 2017 and June. 2018. 
Each data sample consists of the open, closing, lowest, highest price and trading volume in the past time windows and a target trading volume.
We adopt the data of 2017 as the training set and development set, and the samples between Jan. 2018 and
Jun. 2018 are adopted as the test set, making sure that the test set and the training dataset are non-overlapping.
The dataset statistics can be found in Table~\ref{tab:dataset}.

\subsection{Baselines}
We compare our methods to the following baseline models, including:

\noindent\textbf{Moving average methods}, which adopts the averaged last 20-day transaction data as the predictions. We implement (1) Simple Moving Average~(SMA), where the predictions are the averaged trading volume of the last $20$ days at the same time slot, i.e., $\hat x = \frac{1}{T} \sum_{i=1}^{T} x_i$, and (2) Exponential Moving Average~(EMA), which pays more attention to the nearest values, by setting $y_1 = x_1$ and $y_t = \rho x_t + ( 1 - \rho) y_{t-1} $. $y_T$ is adopted as the prediction. We set $\rho =0.04$ following \cite{zhang2021asat}.
% we adopt �� = 0.04 and try to use the 20-da

\noindent\textbf{Teacher-free methods}, which requires no teacher model for training the student. We implement two methods: (1) Min-MSE, where the student minimizes the mean-square error between the prediction and the ground-truth volume, and (2) DeepAR~\cite{salinas2020deepar}, which models the prediction as a conditional probability distribution and maximizes the log-likelihood of the oracle data. 

\noindent\textbf{Distillation methods}, which utilizes the teacher model for enhancing the student model. Specifically, we implement (1) Vanilla KD~\cite{Hinton2015Distilling}, where the mean-square error of the student predictions and teacher predictions and the objective of the Min-MSE method are both optimized, similar to the original knowledge distillation in the classification problem, and (2) Attentive Imitation Loss~(AIL)~\cite{Saputra2019AIL}, where the supervision from imitating the teacher prediction is adaptively adjusted according to the relative correctness of the teacher model.

\subsection{Implementation Details}
Without loss of generality, we adopt the Transformer~\cite{vaswani2017transformer} model as the backbone model due to its powerful modeling ability.
The teacher model consists of $6$ Transformer layer, which contains a self-attention module and a feed-forward network layer. 
We omit the description of the Transformer layer due to the space limit and refer readers to the original Transformer  paper~\cite{vaswani2017transformer} for details.
The student is a much smaller Transformer model with only one layer. 
The number of input, number of hidden units and the output dimension are all set to $200$, and the hidden states are split into $8$ heads for capturing sub-space relations.
The teacher and the student only differ from the layer number. The resulting numbers of model parameters of the student and the teacher model are $0.3$M and $1.5$M, respectively.
To eliminate the influence of model capacity, all the teacher-free models are of the same architecture with the student model in distillation methods.
% , i.e., and with the student model in the following distillation methods.
For a fair comparison, we set the $\lambda_\text{NLL}$ and $\lambda_\text{DKD}$ to $0.5$, which is consistent with the $\alpha$ in Vanilla KD.
We perform grid search over the hyper-parameter $\lambda_\text{Dist}$ and $\lambda_\text{Cosine}$ in $\{1.0, 2.0, 5.0, 10.0\}$ and select the best performing parameters according to the validation performance.
We adopt the Adam~\cite{adamw} optimizer and initialize the learning rate with $0.001$. 
The batch size is set to $32$ and the consistency matrix is computed in the mini-batch. 
We evaluate the model performance on the validation set every $1000$ steps, and test the best model on the test set.
We repeat every experiment with $7$ random seeds and report the averaged results with a GeForce GTX 1080Ti GPU.

\subsection{Main Results}
\begin{table*}[tbh!]
	\centering
	\small 
	\caption{Experimental results on two different time window settings comparing with different baseline models. The best results are shown in \textbf{bold}. * denotes the results are statistically significant compared with the best performing baseline with $p < 0.05$.}
	\label{tab:res}
\resizebox{\columnwidth}{!}{
	\begin{tabular}{@{}l|cccccc@{}}
		\toprule
		\textbf{Dataset}&\multicolumn{3}{c}{\textbf{Hourly}}&\multicolumn{3}{c}{\textbf{Daily}}\\
% 		\toprule
        \midrule
		\textbf{Model}& \textbf{MSE}~($\downarrow$)&\textbf{MAE}~($\downarrow$)&\textbf{ACC}~($\uparrow$)&\textbf{MSE}~($\downarrow$)&\textbf{MAE}~($\downarrow$)&\textbf{ACC}~($\uparrow$) \\ 
		\midrule 
		Teacher Model
		  &0.195 & 0.335& 0.731 & 0.124 & 0.265& 0.665 \\
		\midrule
		20-day SMA
		  &0.249& 0.385 &  0.709 &  0.169& 0.321 &  0.634\\
		  %MSE: 0.195578, RMSE: 0.442242, MAE: 0.343878, ACC: 0.632269 
		20-day EMA
		& 0.288 &  0.413 &  0.696 & 0.196  & 0.344 & 0.632 \\
% 		20-day EMA
% 		& &0.600&0.713&0.427&0.498&0.727\\
% 		Last Time Slot
% 		&1.118&0.742&0.500&0.653&0.602&0.500 \\
% 		12-slot Average
% 		&0.982&0.710&0.630&0.975&0.782&0.44 5\\
% 		12-slot EMA
% 		&0.888&0.668&0.642&0.846&0.718&0.457 \\
% 		20-day and 12-slot Average
% 		&0.689&0.581&0.713&0.377&0.469&0.698 \\
		\midrule
% Number of Parameters 
% 290,807 1L 
% 1,016,807 4L
% 1,500,807 6L 
		Min-MSE  & 0.206 & 0.348 & 0.719 &0.137  &0.284 & 0.630\\  % 普通版训练 
	    DeepAR~\cite{salinas2020deepar}   &0.204 &  0.347 &  0.721 & 0.139  & 0.288 & 0.627\\ 
		Vanilla KD~\cite{Hinton2015Distilling}   &0.200  & 0.342 & 0.725 &0.132  & 0.277& 0.645\\  % MSE 版本的 KD
% 		Bound KD & & & & & & \\ 
		AIL~\cite{Saputra2019AIL} & 0.202  &0.344 &0.722 & 0.134 &0.282 & 0.634\\ 
	    \midrule 
	    DKD  & 0.201 & 0.342 & 0.724 & 0.133 & 0.279 & 0.639 \\
	    ~w/ Dist-CKD   & 0.202 &  0.344 & 0.726 & 0.129 & 0.272& 0.652 \\ 
	    ~w/ Cosine-CKD  & \textbf{0.197}$^*$   &\textbf{0.339}$^*$ & \textbf{0.728}$^*$ & \textbf{0.128}$^*$ & \textbf{0.271}$^*$ &  \textbf{0.656}$^*$ \\ 
	    ~w/ Dist-CKD + Cosine-CKD   & 0.199  & 0.340  & 0.727 & 0.129 & 0.273 & 0.652 \\
		\bottomrule
	\end{tabular}}
\end{table*}
The main results on the TPX500 dataset with two different time-window settings can be found in Table~\ref{tab:res}.
It can be found that:

(1) Na\"ive moving average methods like 20-day SMA achieves high prediction accuracy, even better than DNN-based models on the daily dataset.
This reflects a strong consistency of stock trading volume in time series, i.e., stocks with larger trading volumes in the last $20$ days will also have more active trading in the future. 
However, regarding the absolute prediction error metrics MSE and MAE, averaging methods fall far behind the DNN-based methods, which indicates that the powerful DNNs are capable of capturing more complex data patterns behind the time series data and thus making closer predictions.
(2) Distillation methods like Vanilla KD and AIL consistently outperform methods without a teacher model. 
This validates the effectiveness of KD by transferring the learned knowledge from a larger teacher model into the student model to help improve the student performance.
(3) Our proposed correlation-aware distillation framework achieves the best performance. 
For example, on the hourly dataset, conducting distributional KD with the cosine similarity correlation objective achieves $99.6\%$ prediction accuracy of the teacher model, while reducing the model size by $5\times$. 
It indicates that conducting KD on the distribution level and incorporating the correlation objectives are effective for enhancing the KD effect.
(4) Interestingly, we observe that the angle-wise objective consistently outperforms the distance-wise correlation, which we attribute to the fact the angle correlation is higher-order information, thus is more effective for the student model to gain knowledge.
Besides, combining two correlation objectives together cannot bring further performance gain, indicating that the knowledge in the two correlations can be overlapped to some extent. We leave the exploration towards better incorporating different correlation objectives as future work.
% 
% It can be found that on two different time datasets, the proposed DKD with angle-wise distillation objective acheives the best performance.
\section{Analysis}
% \subsection{Trade-off Between Sample Diversity and Correlation Sampling}
In this section, we conduct experiments for probing the property of our proposed framework, by exploring the interplay between the distillation objectives, investigating the performance gain under low-resource settings and examining the pair-wise relationship with different distillation methods.
\subsection{Interplay between Distributional KD and Correlational KD}
\begin{table*}[tb!]
	\centering
	\small 
	\caption{Experimental results on two settings. The best results are shown in \textbf{bold}.}
	\label{tab:interplay}
	\begin{tabular}{@{}l|cccccc@{}}
		\toprule
		\textbf{Dataset}&\multicolumn{3}{c}{\textbf{Hourly}}&\multicolumn{3}{c}{\textbf{Daily}}\\
% 		\toprule
        \midrule
		\textbf{Model}& \textbf{MSE}~($\downarrow$)&\textbf{MAE}~($\downarrow$)&\textbf{ACC}~($\uparrow$)&\textbf{MSE}~($\downarrow$)&\textbf{MAE}~($\downarrow$)&\textbf{ACC}~($\uparrow$) \\ 
	    \midrule 
		Vanilla KD~\cite{Hinton2015Distilling}   &0.200  & 0.342 & 0.725 &0.132  & 0.277& 0.645\\ 
	    DKD  & 0.201 & 0.342 & 0.724 & 0.133 & 0.279 & 0.639 \\
	    ~w/ Dist-CKD   & 0.202 &  0.344 & 0.726 & 0.129 & 0.272& 0.652 \\ 
	    ~w/ Cosine-CKD  & \textbf{0.197}   &\textbf{0.339} & \textbf{0.728} & \textbf{0.128} & \textbf{0.271} &  \textbf{0.656} \\ 
	    only Dist-CKD & 0.223 &  0.364 & 0.717 & 0.132 & 0.278&  0.647 \\ 
	    only Cosine-CKD & 0.199   & 0.341 & 0.726 & 0.138 & 0.286 &   0.633\\ 
		\bottomrule
	\end{tabular}
\end{table*}

In our framework, there are two types of distillation objectives: individual distributional distillation objective, i.e., the DKD distillation objective, and pair-wise correlational distillation objectives, i.e., Dist-CKD or Cosine-CKD objective.
However, the interplay between these two distillation objectives remains unclear.
To investigate this, we examine the performance without the DKD distillation objective in Eq.~\ref{eq:dkd} by setting the $\lambda_\text{DKD}$ to $0$. The results are listed in Table~\ref{tab:interplay}.
We find that DKD alone performs worse than Vanilla KD, indicating that directly mimicking the distributional level outputs of teacher models is more challenging for the student than minimizing the discrepancy between single trading volume values.
On the other hand, the correlational distillation objective alone, i.e., only Dist-CKD and only Cosine-CKD also under-perform the Vanilla KD baseline, as only learning the relative correlation of output distributions is not sufficient for the student to predict accurately.
It can be found that only when combined with the correctional distillation objectives, distributional knowledge distillation can achieve the best performance. 
These findings suggest that our proposed framework is holistic, where the two types of distillation objectives are complementary to each other to achieve optimal distillation performance.

% original DKD distillation performs worse than the Vanilla KD, which directly learns from the single valued teacher predictions.

\subsection{Correlational Objectives Boost More with Fewer Data}

\begin{figure}[h!]
    \centering
    \caption{The MSE~(left) and MAE~(right) reduction curve over the non-distillation methods with varying dataset sizes of different distillation objective on the hourly dataset. Our distributional correlation-aware distillation boosts the performance more significantly under low-resource settings. Best viewed in color.}
    \includegraphics[width=0.95\linewidth]{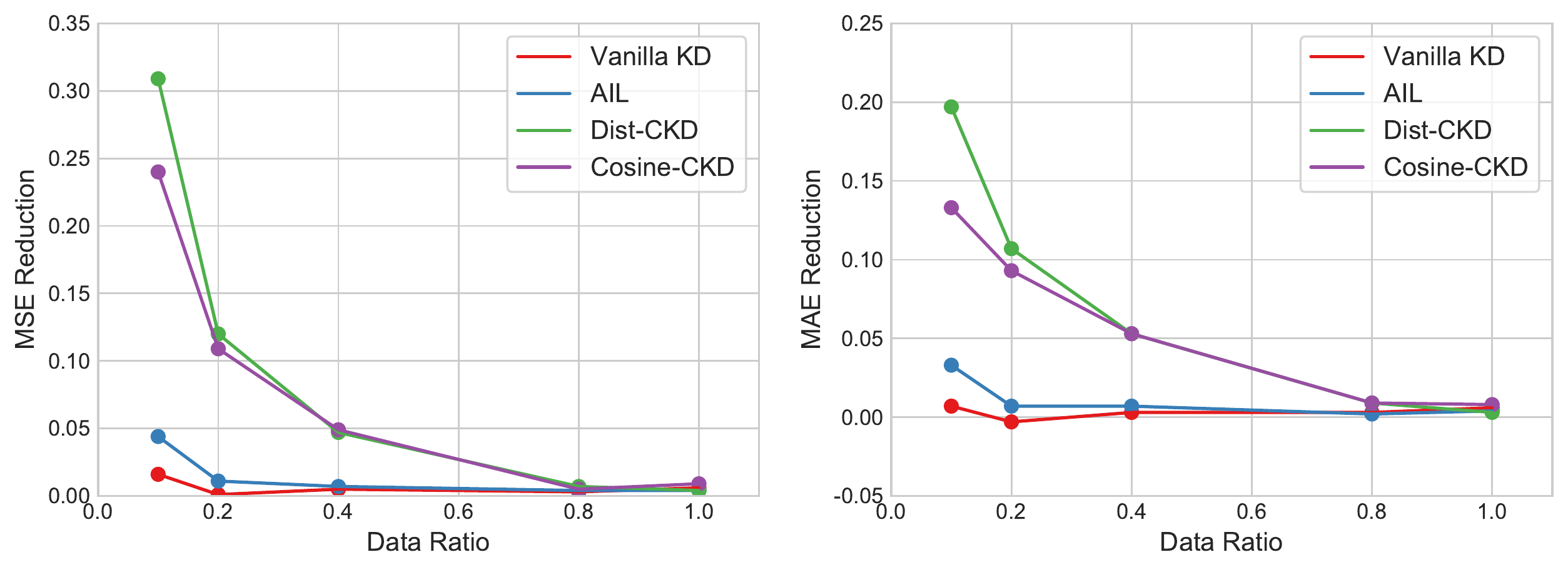}
    \label{fig:data_ratio}
\end{figure}

As our framework can provide more informative supervision than conventional KD, it can be more effective under low-resource settings. To investigate this, we conduct experiments to compare the performance gain over the non-distillation training methods, i.e., Vanilla KD and AIL over Min-MSE and our methods over DeepAR.
% We adopt the searched hyper-parameter settings in our main results in this experiment.
We vary the size of the training dataset from $10\%$ to $100\%$, and plot the performance gain regarding the reduction of the MSE and the MAE with varying training dataset sizes in Figure~\ref{fig:data_ratio}.

Our findings are:
(1) The performance gain of distillation vanishes as the data size becomes larger. 
It is reasonable as the small training dataset cannot provide comprehensive supervision, while the extra information in the teacher predictions can alleviate this problem. 
As the size of the training dataset increases, the training samples cover more diverse data patterns, thus the student can directly learn from the supervision provided by the original data samples instead of relying on the teacher predictions.
This is consistent with previous studies which observe that KD brings more performance boost on small datasets~\cite{Sun2019PatientKD,liang2020mixkd}.
(2) Compared with distillation methods solely based on the predicted scalar values, the proposed Dist-CKD and Cosine-CKD boost the performance more significantly under low-resource settings. We attribute the success to the more informative supervision brought by the distributional correlational-aware distillation objectives, which helps the student make more accurate predictions even with few training samples.
(3) Dist-CKD reduces prediction error more under low-resource settings while Cosine-CKD outperforms the Dist-CKD with the full training dataset. We attribute the reason to that the cosine similarity is a higher-order property than the distance-wise similarity, it may require more data samples to fully exploit its effectiveness.

\subsection{Correlational Objectives Improve Magnitude Ordering}
We are interested in that whether the proposed correlational objective help the model learn the relation between the output trading volumes, which can facilitate better trading decisions.
To investigate this, we calculate the pair-wise relations, i.e., the relative trading magnitude between samples, and probe the relation consistency between the model predictions and the oracle trading volume.
% consistency , i.e., the relative trading volume, between predictions with the oracle volume.
% To investigate this, we devised a measure of relational consistency, i.e. the relative magnitude between model predictions and the consistency of ground truth labels.
Specifically, given $N$ data samples and the corresponding oracle volume $y_1, \ldots, y_N$, we define the error ranking score as:
\begin{equation}
\begin{aligned}
    \text{Error Ranking Number} &= \sum_{i=1}^N \sum_{j=1, j \neq i}^N \text{ErrorSign} \left(  \left(y_i - y_j \right) \times \left( \hat y_i - \hat y_j \right) \right),\\
    \operatorname{ErrorSign} \left( x  \right) &= \begin{cases} 1 & \text { if } x \leq 0 ,\\ 0 & \text { otherwise. }\end{cases}
\end{aligned}
\end{equation}
This metric indicates how many pairs of data samples whose relative trading volume magnitude are mispredicted by the model.
We randomly sampled $12,000$ data samples from the test set and calculate the metric.
The results are shown in Table~\ref{tab:error_number}.
\begin{table*}[t!]
	\centering
	\small 
	\caption{Examining the Error Ranking Number of different methods on datasets with different time window settings. The best results are shown in \textbf{bold}.}
	\label{tab:error_number}
	\begin{tabular}{@{}l|cc@{}}
		\toprule
		\textbf{Dataset}&\textbf{Hourly}&\textbf{Daily}\\
% 		\toprule
        \midrule
		\textbf{Model}& Error Ranking Number~($\downarrow$)& Error Ranking Number~($\downarrow$) \\ 
	    \midrule 
		Min-MSE & 20,545,150 & 14,781,916 \\ 
		Vanilla KD~\cite{Hinton2015Distilling}   & 20,011,092 &14,369,730 \\
		AIL~\cite{Saputra2019AIL} & 20,026,808  &  14,580,932\\ 
		DeepAR~\cite{salinas2020deepar} &20,042,538  &  14,490,082\\ 
		\midrule 
	    DKD  & 20,036,694  & 14,363,378\\ 
	    \quad w/ Dist-CKD & \textbf{19,841,196}  & \textbf{14,294,670}\\
	    \quad w/ Cosine-CKD &  19,930,742 &14,393,534 \\ 
		\bottomrule
	\end{tabular}
\end{table*}
Our observations are:
(1) Distillation methods are effective for improving the prediction consistency with the oracle model. 
Compared with training the student model with the original mean-square error loss objective, Vanilla KD and AIL both greatly reduce the error ranking number. This shows that transferring the knowledge from the teacher model to the student not only improves the student prediction accuracy, but also makes the student become more aware of the relative magnitude between predictions.
(2) Our methods achieve the best performance on both datasets. For example, on the hourly dataset, compared with the Min-MSE method, Dist-CKD reduces the Error Ranking Number by $3.4\%$, which verifies that our proposed correlational distillation objectives can help the model learn the relative trading volume magnitude better.
Besides, Dist-CKD performs consistently better than Cosine-CKD regarding the trading volume magnitude relationship, which we attribute to the fact that the distance correlation defined in Eq.~\ref{eq:dist} is a more explicit modeling of the magnitude relation than the angle-wise objective.

\section{Related Work}
% Our work is mainly related to recent progresses in neural network compression techniques like knowledge distillation, and the explorations of machine learning methods for volume prediction.
\subsection{Knowledge Distillation}
Neural network compression can produce light-weight models for efficient deployments, and it has been an active research area towards green and sustainable deep learning~\cite{xu2021survey,li2021muki,li2021CascadeBERT}. 
% Different from pruning~\cite{han2015deep} and quantization~\cite{hubara2016binaryNN} that require hardware-level support for gaining actual inference speed-up, 
Knowledge distillation~\cite{Hinton2015Distilling,romero2014fitnets} transfers the knowledge of a larger teacher to a smaller student model, achieving a better trade-off between model performance and inference efficiency. 
Recent studies show that KD is effective in computer vision~\cite{park2019relationalKD,Mirzadeh2020TAKD} and natural language processing~\cite{Sun2019PatientKD,Jiao2019TinyBERT}, successfully training a compact student model to perform on par with the large teacher model. 
While previous studies focus on classification problems, in this paper, we explore knowledge distillation for obtaining a compact student model to perform time series forecasting. We build a distributional level knowledge distillation framework for trading volume prediction and propose two correlation-aware distillation objectives.
Our work is partially inspired by \cite{park2019relationalKD}, which aligns the pair-wise correlations of data representations in the teacher and the student model. 
However, our framework focuses on the relationship of output distributions. Their method thus is orthogonal to our method and can be incorporated into our framework for further performance boost. 
Besides, to the best of our knowledge, we are the first to conduct knowledge distillation for the trading volume prediction problem and prove it is effective for obtaining light-weight and well-performing models.

\subsection{Volume Prediction}
Stock trading volume prediction has a significant role in algorithmic
trading systems~\cite{bialkowski2008improving,brownlees2011intra,cartea2016closed}, which aims to predict the stock trading volume based on preceding transaction data. 
Recently, progresses have been made towards more accurate volume prediction via various machine learning techniques.
Specifically, \cite{liu2017intraday} propose to adopt support vector machine~(SVM) for the regression problem to predict the changes of volume percentage. \cite{Libman2019VolumePW} exploit long short-term memory~(LSTM) models~\cite{hochreiter1997lstm} for its capability of modeling long-range dependency. 
Besides, temporal mixture ensemble models~\cite{AntulovFantulin2020TemporalME}, Bayesian auto-regressive models~\cite{Huptas2018PointFO} and graph neural networks~\cite{Zhao2021LongtermSA} are also explored in volume prediction.
\cite{zhang2021asat} train a Transformer model~\cite{vaswani2017transformer} with adversarial objectives to improve the model performance and robustness at the same time.
In this paper, we focus on distilling a more efficient trading volume prediction model and adopt the powerful Transformer as the backbone model for distillation. Our framework is generalizable and can be easily extended to other backbone models.

\section{Conclusion}
In this paper, we present a distributional knowledge distillation framework for training light-weight trading volume prediction models.
The learned knowledge of the teacher model is transferred to the student model at the distributional level, by minimizing the KL-divergence between the predicted Gaussian distributions and the distance-wise and angle-wise correlation distillation objectives.
Experiments on the TPX500 dataset with two different time window settings show that our framework can effectively compress the model while maintaining accurate predictions.
Further analysis shows that the correlational objectives significantly boost the student performance under low-resource settings and make the predictions more consistent with the oracle labels.
In the future, we are hoping to explore this framework for more general regression tasks.
\section*{Acknowledgements}
We thank all the anonymous reviewers for their constructive comments.
This work is supported by a Research Grant from Mizuho Securities
Co., Ltd. We sincerely thank Mizuho Securities for valuable domain
expert suggestions and the experiment dataset. Xu Sun and Ruihan Bao are the corresponding authors of this paper.
\bibliographystyle{splncs04}

\begin{thebibliography}{10}
\providecommand{\url}[1]{\texttt{#1}}
\providecommand{\urlprefix}{URL }
\providecommand{\doi}[1]{https://doi.org/#1}

\bibitem{AntulovFantulin2020TemporalME}
Antulov-Fantulin, N., Guo, T., Lillo, F.: Temporal mixture ensemble models for
  intraday volume forecasting in cryptocurrency exchange markets. arXiv:
  Trading and Market Microstructure  (2020)

\bibitem{bialkowski2008improving}
Bia{\l}kowski, J., Darolles, S., Le~Fol, G.: Improving vwap strategies: A
  dynamic volume approach. Journal of Banking \& Finance  \textbf{32}(9),
  1709--1722 (2008)

\bibitem{brownlees2011intra}
Brownlees, C.T., Cipollini, F., Gallo, G.M.: Intra-daily volume modeling and
  prediction for algorithmic trading. Journal of Financial Econometrics
  \textbf{9}(3),  489--518 (2011)

\bibitem{cartea2016closed}
Cartea, {\'A}., Jaimungal, S.: A closed-form execution strategy to target
  volume weighted average price. SIAM Journal on Financial Mathematics
  \textbf{7}(1),  760--785 (2016)

\bibitem{devlin2019bert}
Devlin, J., Chang, M.W., Lee, K., Toutanova, K.: {BERT}: Pre-training of deep
  bidirectional transformers for language understanding. In: NAACL-HLT. pp.
  4171--4186 (2019)

\bibitem{dosovitskiy2020vit}
Dosovitskiy, A., Beyer, L., Kolesnikov, A., Weissenborn, D., Zhai, X.,
  Unterthiner, T., Dehghani, M., Minderer, M., Heigold, G., Gelly, S., et~al.:
  An image is worth 16x16 words: Transformers for image recognition at scale.
  In: ICLR (2020)

\bibitem{furlanello2018born}
Furlanello, T., Lipton, Z.C., Tschannen, M., Itti, L., Anandkumar, A.:
  Born-again neural networks. In: ICML. Proceedings of Machine Learning
  Research, vol.~80, pp. 1602--1611 (2018)

\bibitem{han2015deep}
Han, S., Mao, H., Dally, W.J.: Deep compression: Compressing deep neural
  networks with pruning, trained quantization and huffman coding. arXiv
  preprint arXiv:1510.00149  (2015)

\bibitem{Hinton2015Distilling}
Hinton, G., Vinyals, O., Dean, J.: Distilling the knowledge in a neural
  network. arXiv preprint arXiv:1503.02531  (2015)

\bibitem{hochreiter1997lstm}
Hochreiter, S., Schmidhuber, J.: Long short-term memory. Neural computation
  \textbf{9}(8),  1735--1780 (1997)

\bibitem{hubara2016binaryNN}
Hubara, I., Courbariaux, M., Soudry, D., El{-}Yaniv, R., Bengio, Y.: Binarized
  neural networks. In: NeurIPS. pp. 4107--4115 (2016)

\bibitem{Huptas2018PointFO}
Huptas, R.: Point forecasting of intraday volume using bayesian autoregressive
  conditional volume models. Journal of Forecasting  (2018)

\bibitem{Jiao2019TinyBERT}
Jiao, X., Yin, Y., Shang, L., Jiang, X., Chen, X., Li, L., Wang, F., Liu, Q.:
  Tinybert: Distilling bert for natural language understanding. In: Findings of
  the Association for Computational Linguistics: EMNLP 2020. pp. 4163--4174
  (2020)

\bibitem{li2021CascadeBERT}
Li, L., Lin, Y., Chen, D., Ren, S., Li, P., Zhou, J., Sun, X.: {C}ascade{BERT}:
  Accelerating inference of pre-trained language models via calibrated complete
  models cascade. In: Findings of the Association for Computational
  Linguistics: EMNLP. pp. 475--486 (2021)

\bibitem{Li2021DynamicKD}
Li, L., Lin, Y., Ren, S., Li, P., Zhou, J., Sun, X.: Dynamic knowledge
  distillation for pre-trained language models. In: EMNLP. pp. 379--389 (2021)

\bibitem{li2021muki}
Li, L., Lin, Y., Ren, X., Zhao, G., Li, P., Zhou, J., Sun, X.: Model
  uncertainty-aware knowledge amalgamation for pre-trained language models.
  arXiv preprint arXiv:2112.07327  (2021)

\bibitem{liang2020mixkd}
Liang, K.J., Hao, W., Shen, D., Zhou, Y., Chen, W., Chen, C., Carin, L.:
  {MixKD}: Towards efficient distillation of large-scale language models. In:
  ICLR (2021)

\bibitem{Libman2019VolumePW}
Libman, D.S., Haber, S., Schaps, M.: Volume prediction with neural networks.
  Frontiers in Artificial Intelligence  \textbf{2} (2019)

\bibitem{liu2017intraday}
Liu, X., Lai, K.K.: Intraday volume percentages forecasting using a dynamic
  svm-based approach. Journal of Systems Science and Complexity
  \textbf{30}(2),  421--433 (2017)

\bibitem{adamw}
Loshchilov, I., Hutter, F.: Decoupled weight decay regularization. In: ICLR
  (2019)

\bibitem{Mirzadeh2020TAKD}
Mirzadeh, S., Farajtabar, M., Li, A., Levine, N., Matsukawa, A., Ghasemzadeh,
  H.: Improved knowledge distillation via teacher assistant. In: AAAI. pp.
  5191--5198 (2020)

\bibitem{pardo2018gaussianKL}
Pardo, L.: Statistical inference based on divergence measures. Chapman and
  Hall/CRC (2018)

\bibitem{park2019relationalKD}
Park, W., Kim, D., Lu, Y., Cho, M.: Relational knowledge distillation. In:
  CVPR. pp. 3967--3976 (2019)

\bibitem{romero2014fitnets}
Romero, A., Ballas, N., Kahou, S.E., Chassang, A., Gatta, C., Bengio, Y.:
  Fitnets: Hints for thin deep nets. In: ICLR (2015)

\bibitem{salinas2020deepar}
Salinas, D., Flunkert, V., Gasthaus, J., Januschowski, T.: Deepar:
  Probabilistic forecasting with autoregressive recurrent networks.
  International Journal of Forecasting  \textbf{36}(3),  1181--1191 (2020)

\bibitem{Sanh2019DistilBERT}
Sanh, V., Debut, L., Chaumond, J., Wolf, T.: Distil{BERT}, a distilled version
  of {BERT}: smaller, faster, cheaper and lighter. In: NeurIPS Workshop on
  Energy Efficient Machine Learning and Cognitive Computing (2019)

\bibitem{Saputra2019AIL}
Saputra, M.R.U., de~Gusm{\~{a}}o, P.P.B., Almalioglu, Y., Markham, A., Trigoni,
  N.: Distilling knowledge from a deep pose regressor network. In: ICCV. pp.
  263--272 (2019)

\bibitem{Shen2020QBERT}
Shen, S., Dong, Z., Ye, J., Ma, L., Yao, Z., Gholami, A., Mahoney, M.W.,
  Keutzer, K.: Q-{BERT}: Hessian based ultra low precision quantization of
  {BERT}. In: AAAI. pp. 8815--8821 (2020)

\bibitem{Sun2019PatientKD}
Sun, S., Cheng, Y., Gan, Z., Liu, J.: Patient knowledge distillation for {BERT}
  model compression. In: EMNLP-IJCNLP. pp. 4323--4332 (2019)

\bibitem{vaswani2017transformer}
Vaswani, A., Shazeer, N., Parmar, N., Uszkoreit, J., Jones, L., Gomez, A.N.,
  Kaiser, L., Polosukhin, I.: Attention is all you need. In: NeurIPS. pp.
  5998--6008 (2017)

\bibitem{xu2021survey}
Xu, J., Zhou, W., Fu, Z., Zhou, H., Li, L.: A survey on green deep learning.
  arXiv preprint arXiv:2111.05193  (2021)

\bibitem{zhang2021asat}
Zhang, Z., Li, W., Bao, R., Harimoto, K., Wu, Y., Sun, X.: {ASAT}: Adaptively
  scaled adversarial training in time series. arXiv preprint arXiv:2108.08976
  (2021)

\bibitem{Zhao2021LongtermSA}
Zhao, L., Li, W., Bao, R., Harimoto, K., Wu, Y., Sun, X.: Long-term, short-term
  and sudden event: Trading volume movement prediction with graph-based
  multi-view modeling. In: Zhou, Z. (ed.) IJCAI. pp. 3764--3770 (2021)

\end{thebibliography}

\appendix 
\section{Cosine Similarity of Gaussian Distributions}
\label{apx:proof}
% We provide the derivation of the inner-dot between Gaussian distributions $\mathcal{N}_i\left(\mu, \sigma \right)$ and  $\mathcal{N}_j\left(\mu_j, \sigma_j \right)$ as below: 
\begin{proof}
The inner-dot and the cosine similarity of $\mathcal{N}_i\left( \mu_i, \sigma_i\right)$ and  $\mathcal{N}_j\left( \mu_j, \sigma_j\right)$ are:
\begin{align*}
(\mathcal{N}_i, \mathcal{N}_j)&=\int_{-\infty}^{+\infty}\mathcal{N}_i(t\mid\mu_i,\sigma_i)\mathcal{N}_j(t\mid\mu_j,\sigma_j)dt\\
&=\int_{-\infty}^{+\infty}\frac{1}{2\pi\sigma_i^2\sigma_j^2}\exp\left(-\frac{(t-\mu_i)^2}{2\sigma_i^2}-\frac{(t-\mu_j)^2}{2\sigma_j^2}\right)dt\\
&=\int_{-\infty}^{+\infty}\frac{1}{2\pi\sigma_i\sigma_j}\exp\left(-\frac{(t-\mu')^2}{2\frac{\sigma_i^2\sigma_j^2}{\sigma_i^2+\sigma_j^2}}-\frac{(\mu_i-\mu_j)^2}{2(\sigma_i^2+\sigma_j^2)}\right)dt\\
&=\frac{1}{\sqrt{2\pi(\sigma_i^2+\sigma_j^2)}}\exp\left(-\frac{(\mu_i-\mu_j)^2}{2(\sigma_i^2+\sigma_j^2)}\right)\\
\end{align*}
where $\mu'=\frac{\mu_i\sigma_j^2+\mu_j\sigma_i^2}{\sigma_i^2+\sigma_j^2}$, $(\mathcal{N}_i, \mathcal{N}_i)=\frac{1}{\sqrt{4\pi\sigma_i^2}}, (\mathcal{N}_j, \mathcal{N}_j)=\frac{1}{\sqrt{4\pi\sigma_j^2}}$,
\begin{align*}
\varphi_\text{Cosine}\left(\mathcal{N}_i, \mathcal{N}_j\right) &=\frac{(\mathcal{N}_i, \mathcal{N}_j)}{\sqrt{(\mathcal{N}_i, \mathcal{N}_i)(\mathcal{N}_j, \mathcal{N}_j)}}\\
&= \frac{\sqrt{(4\pi\sigma_i^2)^{\frac{1}{2}}(4\pi\sigma_j^2)^{\frac{1}{2}}}}{\sqrt{2\pi(\sigma_i^2+\sigma_j^2)}}\exp\left(-\frac{(\mu_i-\mu_j)^2}{2(\sigma_i^2+\sigma_j^2)}\right)\\
&= \sqrt{\frac{2\sigma_i\sigma_j}{{\sigma_i^2 + \sigma_j^2}}}\exp \left( - \frac{\left(\mu_i - \mu_j\right)^2}{2\left(\sigma_i^2 + \sigma_j^2\right)} \right)
\end{align*}
\end{proof}
\end{document}